\documentclass[12pt]{elsarticle}

\usepackage[T1]{fontenc}

\usepackage[utf8]{inputenc} 
\usepackage{amsfonts} 
\usepackage{amssymb}
\usepackage{amsmath}

\usepackage{amsthm}
\theoremstyle{remark}
\newtheorem*{remark}{Remark}

\usepackage{graphicx}
\usepackage{url}
\usepackage{array}
\usepackage{booktabs}
\usepackage{siunitx}
\usepackage{epstopdf}
\usepackage{float}
\usepackage{braket}
\usepackage{tikz}
\usetikzlibrary{calc}
\usetikzlibrary{angles}
\usepackage{pgfplots}
\usepackage{multirow}
\usepackage{setspace}
\usepackage{subcaption}
\usepackage{bm}

\usepackage[breaklinks=true]{hyperref}

\usepackage{color}
\usepackage{mathtools}
\usepackage{tikz}
\usetikzlibrary{shapes,arrows}

\usepackage{nth}

\mathtoolsset{showonlyrefs}

\usepackage{upgreek}

\newcommand{\vect}[1]{%
	{\boldsymbol{\mathbf{%
				\mathit{#1}
	}}}
}

\usepackage{bbm}
\usepackage{bbold} 

\usepackage{fontawesome}

\newcommand\norm[1]{\left\lVert#1\right\rVert}

\usepackage{algpseudocode,algorithm,algorithmicx}

\algrenewcommand\algorithmicrequire{\textbf{Precondition:}}
\algrenewcommand\algorithmicensure{\textbf{Postcondition:}}
\algnewcommand\algto{\textbf{ to }}

\graphicspath{{./figures/}} 
\usepackage{physics}

\usepackage{numprint} 

\usepackage{multirow}

\makeatletter
\newcommand{\printfnsymbol}[1]{%
	\textsuperscript{\@fnsymbol{#1}}%
}
\makeatother

\journal{}

\makeatletter
\def\ps@pprintTitle{%
	\let\@oddhead\@empty
	\let\@evenhead\@empty
	\let\@oddfoot\@empty
	\let\@evenfoot\@oddfoot
}
\makeatother

\begin{document}

\begin{frontmatter}
	
\title{Modular parametric PGD enabling online solution of partial differential equations}

\author[aff1,aff2]{Angelo Pasquale\corref{cor1}}
\cortext[cor1]{Corresponding author}
\ead{angelo.pasquale@ensam.eu}
\author[aff2]{Mohammad-Javad Kazemzadeh-Parsi}
\ead{javad.kazem@ensam.eu}
\author[aff1,aff3]{Daniele Di Lorenzo}
\ead{daniele.di\_lorenzo@ensam.eu}
\author[aff1]{Victor Champaney}
\ead{victor.champaney@ensam.eu}
\author[aff2,aff4]{Amine Ammar}
\ead{amine.ammar@ensam.eu}
\author[aff1,aff4]{Francisco Chinesta}
\ead{francisco.chinesta@ensam.eu}

\affiliation[aff1]{organization={ESI Group Chair @ PIMM Lab, Arts et M\'etiers Institute of Technology},
	addressline={151 Boulevard de l'H\^opital},
	city={Paris},
	postcode={F-75013},
	country={France}}
\affiliation[aff2]{organization={LAMPA Lab, Arts et M\'etiers Institute of Technology},
	addressline={2 Boulevard du Ronceray BP 93525},
	city={Angers cedex 01},
	postcode={49035},
	country={France}}
\affiliation[aff3]{organization={ESI Group},
	addressline={3 bis Saarinen, Parc Icade, Immeuble le Seville},
	city={Rungis Cedex},
	postcode={94528},
	country={France}}
\affiliation[aff4]{organization={CNRS@CREATE LTD},
	addressline={1 Create Way, \#08-01 CREATE Tower},
	city={Singapore},
	postcode={138602},
	country={Singapore}}

\begin{abstract}
	In the present work, a new methodology is proposed for building surrogate parametric models of engineering systems based on modular assembly of pre-solved modules. Each module is a generic parametric solution considering parametric geometry, material and boundary conditions. By assembling these modules and satisfying continuity constraints at the interfaces, a parametric surrogate model of the full problem can be obtained. In the present paper, the PGD technique in connection with NURBS geometry representation is used to create a parametric model for each module. In this technique, the NURBS objects allow to map the governing boundary value problem from a parametric non-regular domain into a regular reference domain and the PGD is used to create a reduced model in the reference domain. In the assembly stage, an optimization problem is solved to satisfy the continuity constraints at the interfaces. The proposed procedure is based on the offline--online paradigm: the offline stage consists of creating multiple pre-solved modules which can be afterwards assembled in almost real-time during the online stage, enabling quick evaluations of the full system response. To show the potential of the proposed approach some numerical examples in heat conduction and structural plates under bending are presented.
\end{abstract}


\end{frontmatter}

\tableofcontents


%

\section{Introduction}
\label{sec:intro}

Within the framework of reduced-order models (ROMs), numerous intrusive and non-intrusive techniques have been developed to solve parametrized partial differential equations (pPDEs), for a large variety of engineering applications \cite{MOR-book-1, MOR-snapshots, MOR-applications}.

Usual techniques are the snapshots-based parametric ROMs (pROMs), which rely on an offline stage where the parametric space is explored computing high-fidelity solutions of the PDE for sampled combinations of parameters values, called sampling points. This stage basically consists of exploring the so-called solution manifold for extracting a reduced approximation basis. 

Within the context of the reduced basis method (RBM) \cite{rb-book, rb-book-1, rb-article, rb-chapter}, the online stage consists of projecting the solution of the full-order model (FOM) over the previously extracted reduced basis and solving the reduced problem. 

Other approaches proceed by directly interpolating among the sampled snapshots, extracted orthogonal bases or subspaces \cite{SSL,sPGD, sPGD-and-variants,sPGD-and-variants-1,PODI,parametric-model-HMC,parametric-battery,parametric-battery-conference}. This accelerates and simplifies the procedure, at the expense of loosing accuracy since no reduced problem is solved during the online stage. In fact, in many cases, to ensure robustness with respect to parameters variations, the interpolation must be performed on the solution manifold \cite{Grassmanian-interpolation, Grassmanian-interpolation-1, Grassmanian-interpolation-1a, Grassmanian-interpolation-2}. Otherwise, recent studies suggest new interpolation strategies, such as parametric optimal transport \cite{OT}.

Another family of approaches is the one coming from the proper generalized decomposition (PGD) \cite{pgd-article, pgd-book}, where a pPDE is solved accounting for the parameters as extra-coordinates, additionally to usual space and time coordinates. The offline stage, in this context, consists of solving a high-dimensional problem exploiting the separation of variables and defining a fixed-point based iterative algorithm. Recent studies combine the PGD-based parametric solver with NURBS-based geometrical descriptions, allowing to solve efficiently geometrically parametrized PDEs \cite{iga-pgd-1,iga-pgd-3}.
As a main disadvantage, due to its intrusiveness, the PGD procedure often requires ad hoc implementations which can be difficult in case of large-scale problems or in industrial settings.

All the previously mentioned works share common issues when dealing with large and complex systems, where a single simulation can be excessively expensive computationally. Here, the curse of dimensionality is encountered since such systems often exhibit a high number of parameters, compromising a rich exploration of the parametric space.

To overcome, or at least alleviate, such drawbacks, localized model reduction methods have been proposed, where standard model reduction is combined with multiscale (MS) or domain decomposition (DD) techniques \cite{MOR-snapshots,DD-ROM-review}. The primary concept of localized model reduction is the decomposition of the computational domain in modules and the definition of local reduced models, which are coupled across interfaces.

A first example is the usage of finite element tearing and interconnecting (FETI) \cite{far91,far92,far94}, where the computational domain is divided into smaller subdomains or substructures, and then interconnected using interface degrees of freedom. The compatibility is enforced efficiently regardless of the differences in meshing strategies, making the method suitable for large-scale structures composed of multiple components or materials.
	
In the context of RBM, one can refer to \cite{DD-ROM-RBM-0, DD-ROM-RBM, DD-ROM-RBM-1}, where authors target many-parameter thermo-mechanical analyses over repeated components systems. Moreover, in \cite{static-condensation}, the static condensation reduced basis method has recently been applied to efficiently model parametric wind turbines. The method has also been applied in \cite{static-condensation-1} to model general cellular structures. In \cite{DD-ROM-GPOD, DD-ROM-POD}, several ROMs based on the proper orthogonal decomposition (POD) have been proposed for fluid-dynamics and neutron diffusion problems. Other studies successfully combine domain partitioning strategies with projection-based ROMs and hyper-reduction approaches in nonlinear and chaotic fluid-dynamics settings \cite{DD-ROM-PROJ-local-bases, DD-ROM-PROJ-cfd,DD-ROM-PROJ-turbulent-0, DD-ROM-PROJ-turbulent}. 

Recent works suggest nonlinear manifold ROMs based on modules modeling via neural-networks, sparse autoencoders and hyper-reduction \cite{DD-ROM-NN}. Also in the framework of physics-informed neural networks, domain decomposition is introduced to tackle large-scale problems \cite{DD-PINN}. 

In some works, the interface coupling benefits of standard algorithms developed in the literature of DD. For instance, in \cite{DD-ROM-Schwarz} the Schwarz alternating method is used to enable ROM-FOM and ROM-ROM coupling in nonlinear solid mechanics. Similarly, in \cite{DD-ROM-Schwarz-overl} the one-shot overlapping Schwarz approach is applied to component-based MOR of steady nonlinear PDEs. In \cite{DD-ROM-Schur, DD-ROM-Schur-1} the transmission problem along the interface is formulated in terms of a Lagrange multiplier representing the interface flux and solved through a dual Schur complement. 

In the context of the proper generalized decomposition, subdomain approaches have been proposed in \cite{iga-pgd-2, DD-PGD, DD-PGD-1, DD-PGD-2}. In \cite{iga-pgd-2} a multi-patch NURBS-PGD approach has been developed with the aim of enabling or simplifying the PGD solution of problems defined over complex domains. In \cite{DD-PGD}, the Arlequin method constructs local PGD solutions and uses Lagrange multipliers in overlapping regions to connect the local surrogates. A non-overlapping Dirichlet–Dirichlet method is instead used in \cite{DD-PGD-1}, where the local surrogates are computed in the offline phase, while an interface problem is solved online to ensure the coupling. In \cite{DD-PGD-2}, a DD-PGD framework is introduced for linear elliptic PDEs, utilizing an overlapping Schwarz algorithm to connect local surrogate models exhibiting parametric Dirichlet boundary conditions.

In the present work, a general component-based pMOR framework (valuable for intrusive and non-intrusive surrogates) is presented. This is based on decomposing the spatial region in non-overlapping parametric patches. Single-patch parametric solutions are built exploiting, without loss of generality, the NURBS-PGD technique (other surrogate modeling choices could be done over a patch, also snapshots-based such as PODI and sPGD \cite{sPGD,sPGD-and-variants,sPGD-and-variants-1}), accounting additionally for parametric boundary conditions, as well as other parameters related to loading or physics. The full-system parametric solution is then built ensuring the equilibrium of patches across the interfaces, that could make ouse of any coupling procedure.

The paper is structure as follows. Section \ref{sec:methods} is describing the overall procedure. Section \ref{sec:results} shows applications two benchmark problems in computational mechanics. Section \ref{sec:results} gives conclusive remarks and perspectives.
 
\section{Materials and methods}
\label{sec:methods}

Let us consider a parametric problem $\mathcal{P}$ (i.e., a pPDE) defined over the physical space $\Omega \subset \mathbb{R}^d$, with $d = 2, 3$, and involving $N_p$ parameters collected in the vector $\boldsymbol{p}$. Moreover, problem $\mathcal{P}$ is equipped of suitable boundary conditions on the boundary of the domain $\partial \Omega$. 

\begin{figure}[h]
	\centering	
	\includegraphics[scale=0.4]{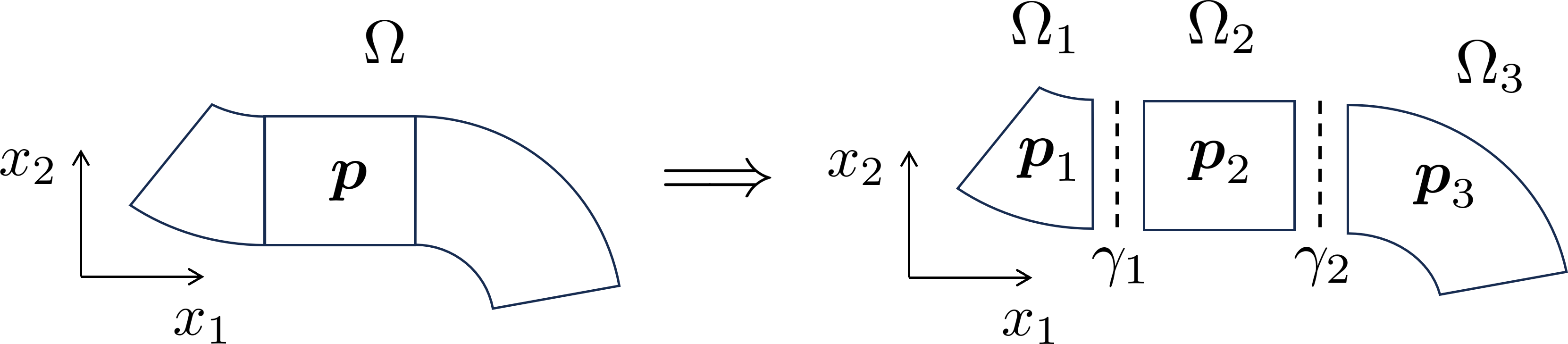}
	\caption{Modularisation of the parametric problem $\mathcal{P}$ defined on domain $\Omega$ and parameter vector $\boldsymbol{p}$ into a set of some pre-solved sub-problems $\mathcal{P}_i$ defined on $\Omega_i$ considering parameter vector $\boldsymbol{p}_i$.}
	\label{fig:domain-decomposition}
\end{figure}

For the sake of illustration, but without loss of generality, let us initially suppose that $\Omega \ \subset \mathbb{R}^2$ (the physical coordinates are denoted with $\boldsymbol{x} = (x_1, x_2)$) is decomposable in a number of non-overlapping parametric domains (three in our illustration), which can be seen as a macro-element (each part may have its own model and/or geometric parameters). This means that $\Omega$ is expressed as $\Omega = \bigcup_{i = 1}^3 \Omega_i$ and the modules intersect only on their interface, that means $\Omega_i \cap \Omega_j = \gamma_{i}$ or $\Omega_i \cap \Omega_j = \emptyset$ based on whether or not the modules are contiguous, as illustrated in figure \ref{fig:domain-decomposition}. For instance, for the illustrated example, only two interfaces $\gamma_1$ and $\gamma_2$ exist.

The initial problem $\mathcal{P}$ involving $N_p$ parameters is recast in terms of three sub-problems $\mathcal{P}_i$ defined over the modules $\Omega_i$ and involving $N_{p,i} < N_p$ parameters, respectively.

\begin{figure}[h]
	\centering	
	\includegraphics[scale=0.4]{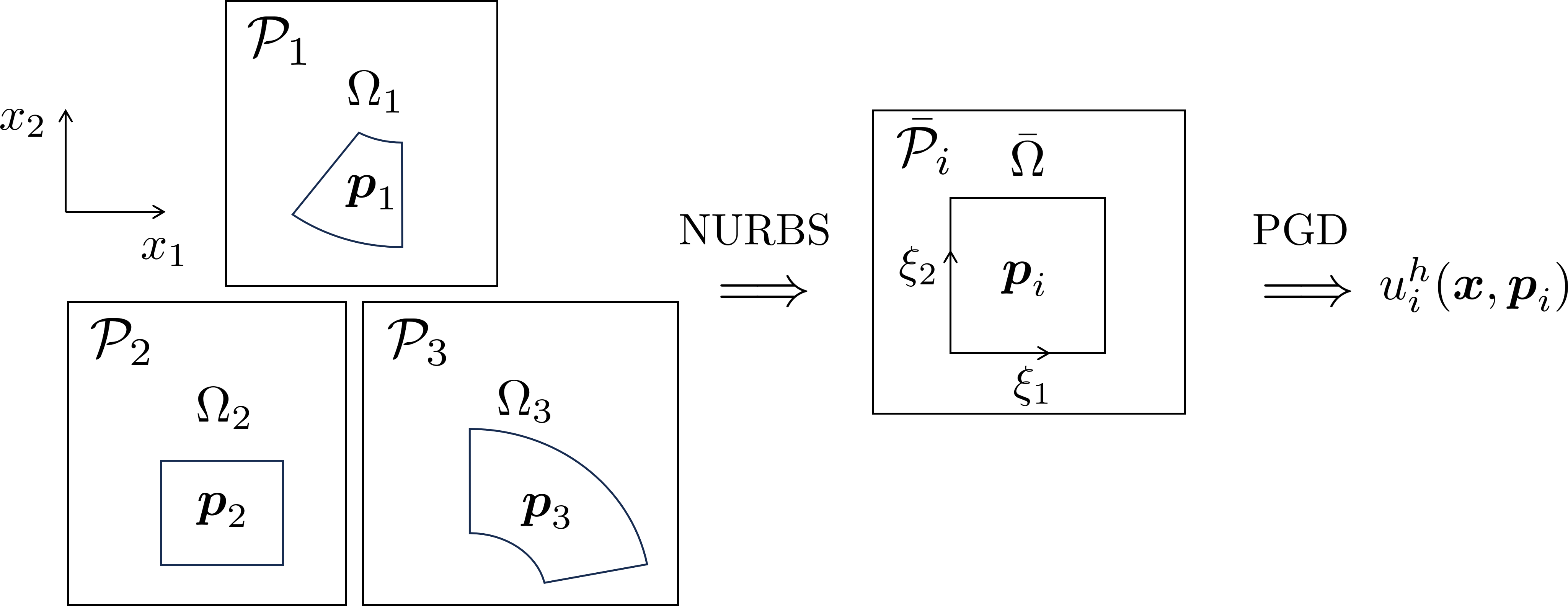}
	\caption{Using the NURBS-PGD technique to map the parametric modules (sub-problems) into a regular (hyper-cubic) computational space and then solving it.}
	\label{fig:subproblems}
\end{figure}

Many possible methods have been discussed in the introduction to address the parametric sub-problems. This study will employ the NURBS-PGD approach, extensively detailed in our previous works, across a wide range of engineering applications, encompassing three-dimensional problems on non-simply connected domains \cite{iga-pgd-1, iga-pgd-3, iga-pgd-2}. This mostly consists of two steps: (a) a NURBS-based geometry mapping from $\Omega$ to the reference square $\bar \Omega$ (the reference coordinates are denoted with $\boldsymbol{\xi} = (\xi_1, \xi_2)$) and (b) the PGD-based computation of a parametric solution $u_i^h(\boldsymbol{\xi}, \boldsymbol{p}_i)$ (and, consequently, $u_i^h(\boldsymbol{x}, \boldsymbol{p}_i)$) related to the local problem $\mathcal{P}_i$, characterized by the parameters collected into the vector $\boldsymbol{p}_i$. The two steps are schematically summarized in figure \ref{fig:subproblems}.

A less direct task is the reconstruction of the parametric solution $u^h(\boldsymbol{x}, \boldsymbol{p})$ of $\mathcal{P}$ starting from the local parametric solutions $u_i^h(\boldsymbol{x}, \boldsymbol{p}_i)$ of $\mathcal{P}_i$. Indeed, the parametric solutions  on the different modules $\Omega_i$ must be assembled across the internal interfaces to reconstruct the response over the whole domain $\Omega$. 

The modules $\Omega_i$ have internal boundaries defined from the common interfaces as $\Gamma_1 = \gamma_1$, $\Gamma_2 = \gamma_1 \cup \gamma_2$ and $\Gamma_3 = \gamma_2$, respectively. Each subproblem $\mathcal{P}_i$ inherits from $\mathcal{P}$ the equations and the imposed boundary conditions over $\partial \Omega \setminus \Gamma_i$. Moreover, each $\mathcal{P}_i$ needs to be equipped of suitable interface conditions (or transmission conditions) over $\Gamma_i$ in order to satisfy the global problem $\mathcal{P}$. 

In the parametric context, the interface conditions must be taken into account within the parametric sub-models. Indeed, the global solution is obtained by the particularization of local solutions at parameters' values, followed by the interfaces equilibrium. To tackle this point, the PGD has the chief advantage to solve BVPs with parametric boundary conditions, treating them as problem extra-coordinates. Following this rationale, the local parametric solutions will be expressed as $u_i^h(\boldsymbol{x}, \boldsymbol{p}_i, \boldsymbol{\alpha}_i)$, where the vector $\boldsymbol{\alpha}_i$ accounts for the boundary conditions imposed over the patch $\Omega_i$, namely the interface conditions.

\begin{figure}[h]
	\centering	
	\includegraphics[scale=0.4]{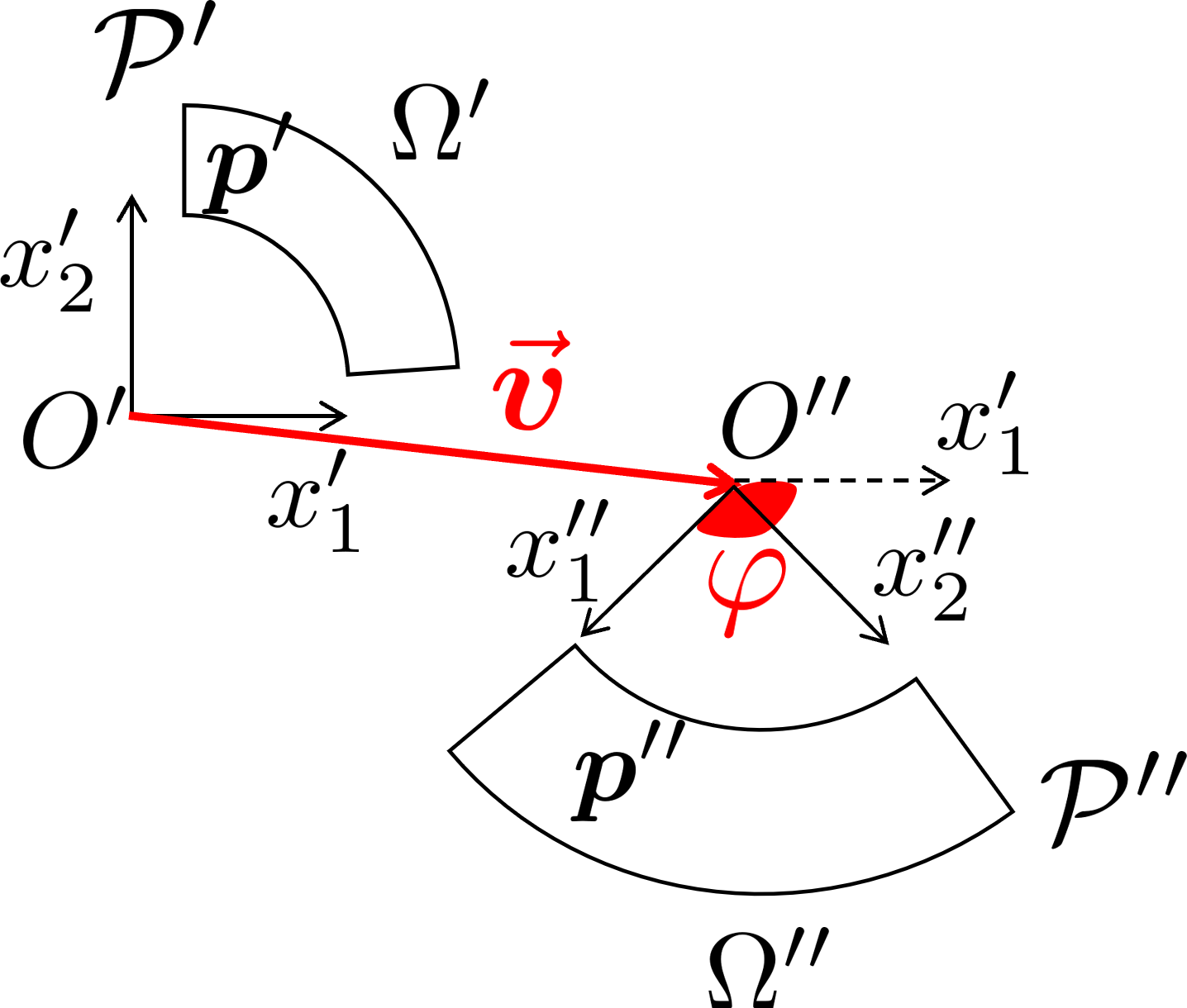}
	\caption{Two modules sharing topology and parameters.}
	\label{fig:replicability}
\end{figure}

As an additional remark, in case of modules sharing topology and parameters, a single parametric sub-model is built and replicated in the assembly of the full system. For instance, as shown in figure \ref{fig:replicability}, the module $\Omega''$ can be obtained from $\Omega'$ defining local frames and opportune translations, rotations and reflections. This is what occurs, for instance, in the case of $\Omega_1$ and $\Omega_3$, which are modeled through a single reference module.

All the details about the parametric modularization and assembly proposed in this work will be given in subsection \ref{subsec:pDD}.

\subsection{Multi-parametric modularization}
\label{subsec:pDD}

Let us consider the general case of a parametric problem $\mathcal{P}$ rewritten in terms of $P$ parametric sub-problems $\mathcal{P}_p$.

The methodology can be summarized with the following steps:
\begin{enumerate}
	\item define the interfaces skeleton by modules decomposition;
	\item identify the model parameters of each module including the parameters considering the geometry, material, model and boundary conditions;
	\item assume a reduced model of the interface conditions (Neumann or Dirichlet);
	\item find a parametric solution for each module, including a parametrization of the interface condition and the model/geometrical parameters;
	\item impose the equilibrium among the modules to determine the global solution.
\end{enumerate}
These steps will be explained in detail in the subsections here below.

\paragraph{Interfaces skeleton}

\begin{figure}[h]
	\centering	
	\includegraphics[scale=0.4]{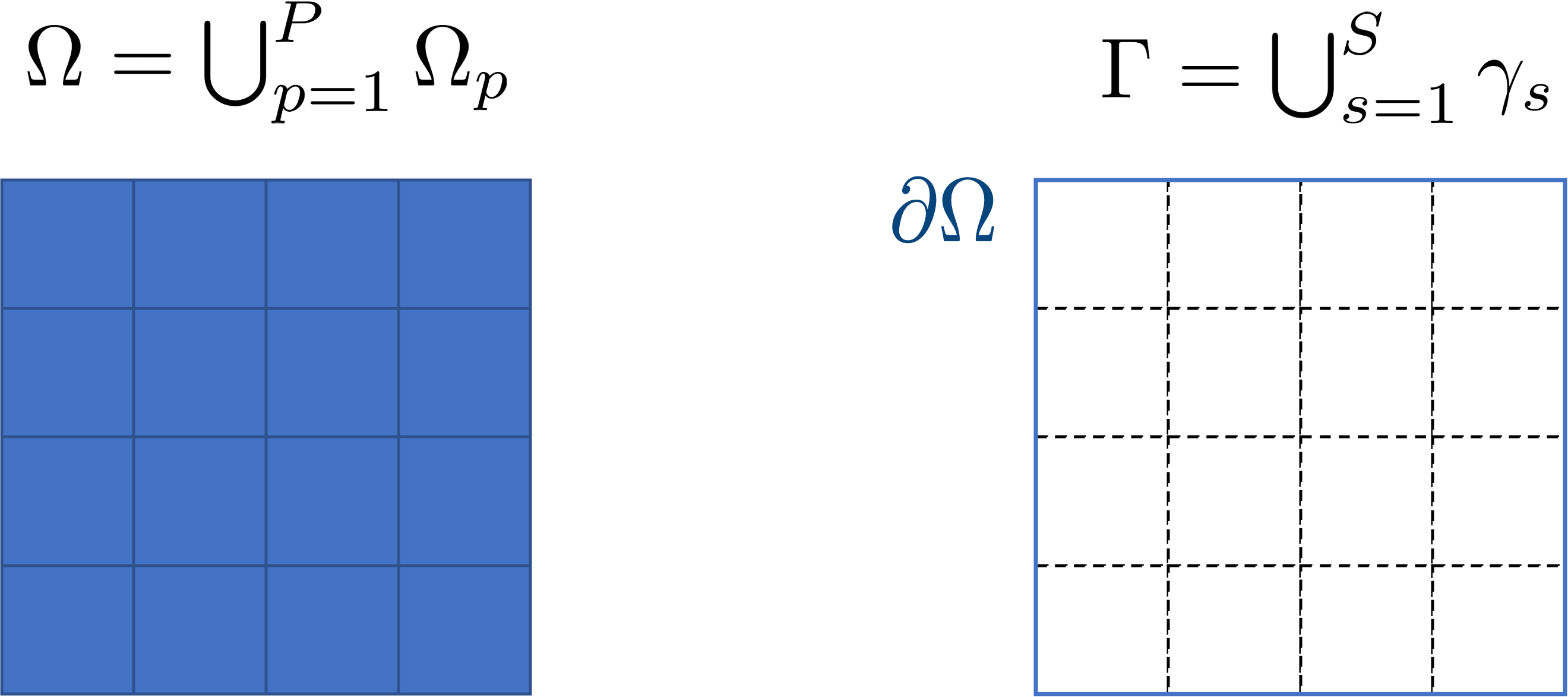}
	\caption{Large structure composed by ${} P$ modules linked by $S$ interfaces.}
	\label{fig:skeleton}
\end{figure}

The domain $\Omega$ is decomposed in $P$ non-overlapping modules $\Omega_p$ and in $S$ internal interfaces $\gamma_s$, $s = 1, \dots, S$, linking the different components and characterizing the structure skeleton $\Gamma$. This is sketched in figure \ref{fig:skeleton}, considering modules as squares. However, as explained previously, any complex shape can be mapped into the square using single-patch or multi-patch NURBS \cite{iga-pgd-3}.

Let us denote with $\mathcal{S}_p$ the set of indices associated with the internal interfaces characterizing $\Omega_p$. This means that $\partial \Omega_p = \Gamma_p \cup \bigcup_{s \in \mathcal{S}_p} \gamma_s$, where $\Gamma_p = \partial \Omega \cap \partial \Omega_p$ is the external part of the boundary.

\begin{figure}[h]
	\centering	
	\includegraphics[scale=0.4]{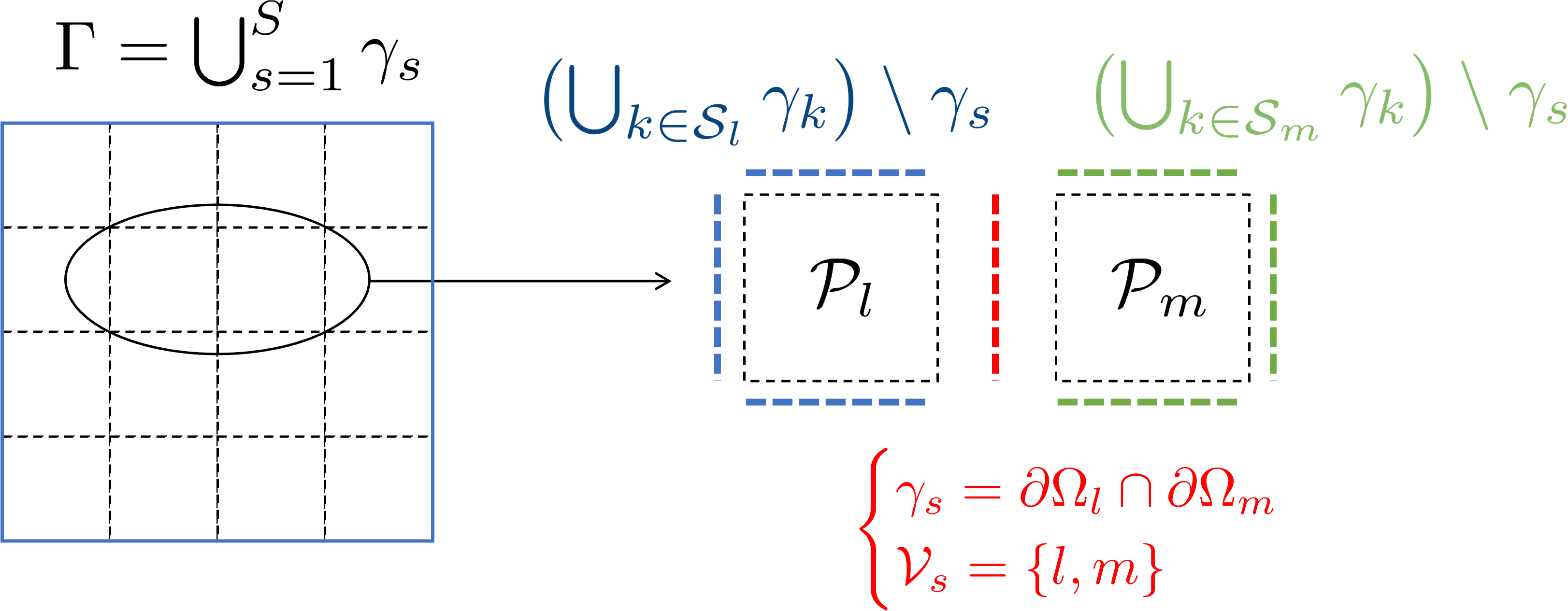}
	\caption{Two modules sharing the same interface $\gamma_s$.}
	\label{fig:common_interface}
\end{figure}

Moreover, let $\mathcal{V}_s$ the set of indices of modules sharing the interface $\gamma_s$. That is, if $\gamma_s = \partial \Omega_l \cap \partial \Omega_m$ then $\mathcal{V}_s = \{l,m\}$. Without loss of generality, this is sketched in figure \ref{fig:common_interface}.

\paragraph{Single part parameters}

Each part composing the entire structure has its own parameters, collected into the vector $\boldsymbol{p}_p$. The parametric sub-problem $\mathcal{P}_p$ is the restriction of the global problem $\mathcal{P}$ to $\Omega_p$. However, this must be equipped of suitable boundary conditions. 

To this purpose, let us suppose that the kinematics at its boundary $\gamma_s$, with $s \in \mathcal{S}_p$, can be reduced to a set of coefficients collected in the vector $\boldsymbol{\alpha}_s$. This means that the whole skeleton reduced kinematics is obtained considering all the internal interface coefficients, that is $\boldsymbol{\Lambda} = (\boldsymbol{\alpha}_1, \dots, \boldsymbol{\alpha}_S)$.

\begin{figure}[h]
	\centering	
	\includegraphics[scale=0.4]{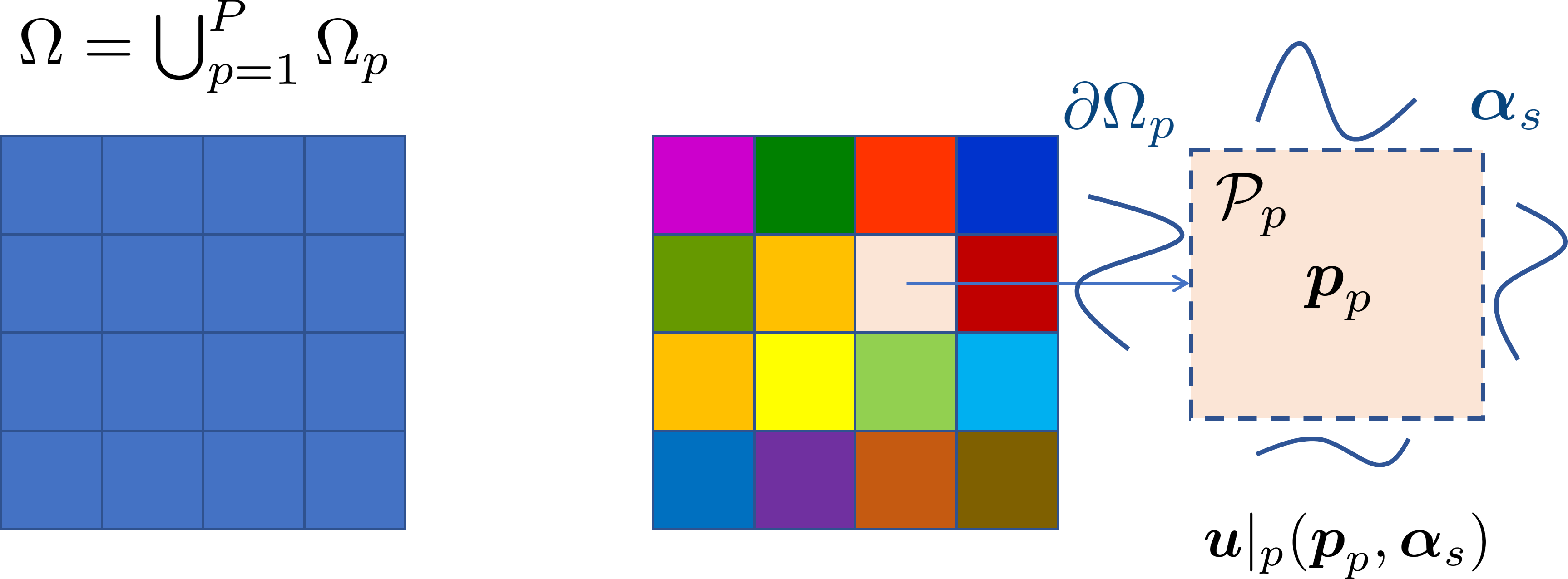}
	\caption{Parametric sub-components including parametric boundary conditions.}
	\label{fig:parametric_lego_alphas}
\end{figure}

Figure \ref{fig:parametric_lego_alphas} shows the definition of a parametric module.

\paragraph{Part reduced model: parametric transfer function}

Each sub-problem $\mathcal{P}_p$ can be solved using the NURBS-PGD method (among other metamodelling strategies), where the parameters $\boldsymbol{p}_p$ and the reduced skeleton kinematics $\boldsymbol{\Lambda}$ are treated as additional coordinates. The parametric transfer function allows the computation of the related thermo-mechanical fields, like the temperature/displacement contour
\begin{equation}\label{PTF}
	\boldsymbol{u}|_{p} = \boldsymbol{u}|_{p} (\boldsymbol{p}_p, \boldsymbol{\alpha}_s), \quad \forall s \in \mathcal{S}_p.
\end{equation} 
In particular, the fluxes/forces at the interface $\gamma_s$ characterizing the part $\Omega_p$ can be extracted
\begin{equation*}
	\boldsymbol{F}|_{s}^p = \boldsymbol{F}|_{s}^p (\boldsymbol{p}_p, \boldsymbol{\alpha}_s), \quad \forall s \in \mathcal{S}_p.
\end{equation*}

\paragraph{Enforcing the interface equilibrium}

The skeleton kinematics $\boldsymbol{u}|_{\Gamma}$ is univocally determined by its reduced representation $\boldsymbol{\Lambda}$. For a new choice of parameters $\boldsymbol{p}_p$, the compatibility of interface fields must be enforced to determine the correct kinematics $\boldsymbol{\Lambda}^\star$. This amounts to minimize a given cost function $\mathcal{C}|_{\Gamma}$ which depends upon the problem at hand.

In structural mechanics, an example of cost function $\mathcal{C}|_{\Gamma}$ can be the sum of forces (and momentums if needed) at all the interfaces $\gamma_s$ (in a thermal problem, this can be the equilibrium of fluxes), that is
\begin{equation*}
	\sum_{s = 1}^{S} \mathcal{C}|_{\gamma_s} = \sum_{s = 1}^{S} \sum_{p \in \mathcal{V}_s} \boldsymbol{F}|_{s}^p = \sum_{s = 1}^{S} \sum_{p \in \mathcal{V}_s} \boldsymbol{F}|_{s}^p (\boldsymbol{p}_p, \boldsymbol{\alpha}_s).
\end{equation*}

\begin{figure}[h!]
	\centering	
	\includegraphics[scale=0.4]{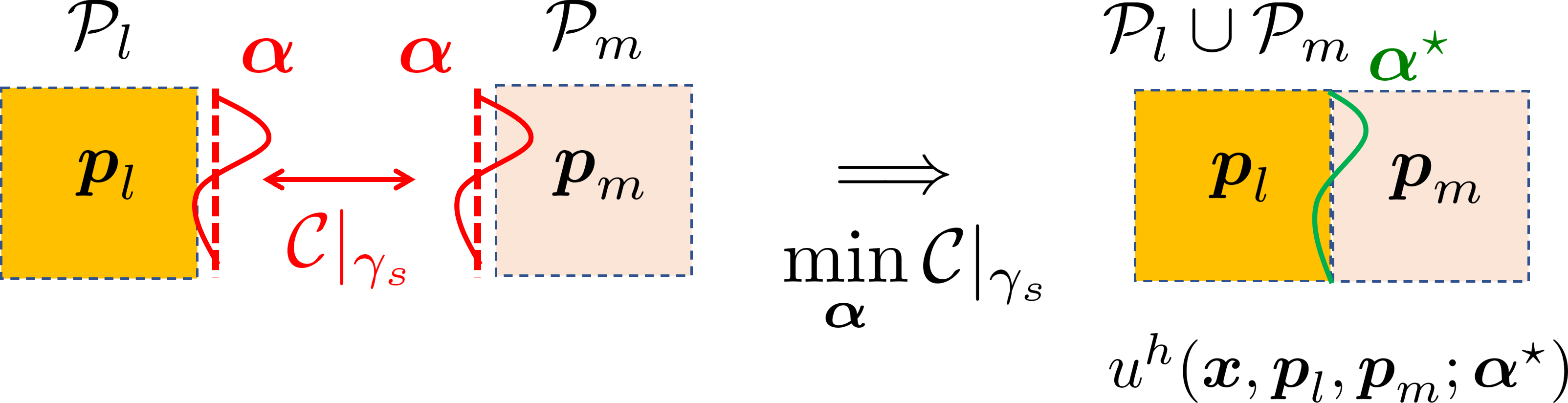}
	\caption{Compatibility of interface between two modules.}
	\label{fig:skeleton_equilibrium}
\end{figure}

For instance, considering two modules $\mathcal{P}_l$ and $\mathcal{P}_m$ sharing the interface $\gamma_s$, this stage consists of determining the coefficients $\boldsymbol{\alpha}^\star$ minimizing the cost function $\mathcal{C}$ at the interface $\gamma_s$. This is illustrated in figure \ref{fig:skeleton_equilibrium}.

\subparagraph{Iterative scheme}

For an arbitrary choice of the skeleton kinematics $\vect{\Lambda}_0 = (\boldsymbol{\alpha}_1^{0},\dots, \boldsymbol{\alpha}_{S}^{0})$, certainly the equilibrium is not satisfied at all the interfaces $\gamma_s$. Thus, we assume the existence of at least one interface violating the equilibrium, e.g. $\gamma_s$

\begin{equation}
	\sum_{p \in \mathcal V_s} \mathbf F |_s^p (\vect{p}_p, \vect{\Lambda}_0) \neq \mathbf 0.
\end{equation}

In what follows the unbalanced forces at each interface will be noted $\mathbf R_s$,
\begin{equation}
	\mathbf R_s = \sum_{p \in \mathcal V_s} \mathbf F |_s^p (\vect{p}_p, \vect{\Lambda}_0), \quad \forall s = 1, \dots, S.
\end{equation}

In order to equilibrate the system one should modify the skeleton kinematics, $\vect{\Lambda}_0 + \Delta \vect{\Lambda}$ (that is, $\boldsymbol{\alpha}_s^{0} + \Delta \boldsymbol{\alpha}_s, \quad \forall s = 1, \dots, S$), in order to satisfy the equilibrium everywhere, we should enforce
\begin{equation}
	\sum_{p \in \mathcal V_s} \mathbf F |_s^p (\vect{p}_p, \vect{\Lambda}_0 + \Delta \vect{\Lambda}) = \mathbf 0, \quad \forall s = 1, \dots, S
\end{equation}
with
\begin{equation}
	\mathbf F |_s^p (\vect{p}_p, \vect{\Lambda}_0 + \Delta \vect{\Lambda}) \approx \mathbf F |_s^p (\vect{p}_p, \vect{\Lambda}_0)+ \left . \frac{\partial \mathbf F |_s^p (\vect{p}_p, \vect{\Lambda})}{\partial \vect{\Lambda}} \right |_{\vect{\Lambda}_0} \Delta \vect{\Lambda},
\end{equation}
leading to the Newton-Raphson iterate
\begin{equation}
	\sum_{p \in \mathcal V_s}  \left . \frac{\partial \mathbf F |_s^p (\vect{p}_p, \vect{\Lambda})}{\partial \vect{\Lambda}} \right |_{\vect{\Lambda}_0} \Delta \vect{\Lambda} = - \mathbf R_s,
\end{equation}
that can be assembled in a linear system
\begin{small}
	\begin{equation}
		\label{sa}
		\left ( 
		\begin{array}{cccc}
			\mathbf M_{11}(\vect{p}_{11}) & \mathbf M_{11}(\vect{p}_{12}) & \cdots & \mathbf M_{1S}(\vect{p}_{1S}) \\
			\mathbf M_{21}(\vect{p}_{21}) & \mathbf M_{21}(\vect{p}_{22}) & \cdots & \mathbf M_{2S}(\vect{p}_{2S}) \\
			\vdots & \vdots & \ddots & \vdots \\
			\mathbf M_{S1}(\vect{p}_{S1}) & \mathbf M_{S1}(\vect{p}_{S 2}) & \cdots & \mathbf M_{SS}(\vect{p}_{SS}) \\
		\end{array}
		\right ) \left (
		\begin{array}{c}
			\Delta \vect{\alpha}_1 \\
			\Delta \vect{\alpha}_2 \\
			\vdots \\
			\Delta \vect{\alpha}_S
		\end{array} 
		\right ) = -
		\left (
		\begin{array}{c}
			\mathbf R_1 \\
			\mathbf R_2 \\
			\vdots \\
			\mathbf R_S
		\end{array} 
		\right ) ,
	\end{equation}
\end{small}
where $\mathbf M_{ij}$ contains the contributions of interface $\gamma_j$ on interface $\gamma_i$, and consequently the parameters $\vect{p}_{ij}$ involved are the ones related to the part that involves interfaces $\gamma_i$ and $\gamma_j$, with $\mathbf M_{ij}=\mathbf 0$ if no part contains both interfaces $\gamma_i$ and $\gamma_j$.

\begin{remark}
	System \eqref{sa} only involves few hundred of equations and consequently can be solved extremely fast. However, in case of many interfaces, its assembly requires the evaluation of the cost function and computation of the gradient (with respect to the parameters), in an iterative setting. Thus, if solved online, the real-time response in some scenarios could be compromised (this is the case for any optimization in high dimension).
	
	A valuable route consists of solving thousands of times (offline) the system \eqref{sa} for a diversity of choices of parameters $\vect{p}_1, \dots , \vect{p}_{P}$, for computing the associated kinematics $\vect{\Lambda} (\vect{p}_1 , \dots , \vect{p}_{P})$. For that purpose powerful regression techniques could be employed, e.g. neural networks-based deep learning.
	
	Another valuable route could be representing the different values of $\vect{\Lambda}_k$, related to the parameter choice $\vect{p}_1^k, \dots , \vect{p}_P^k$, to check its intrinsic dimensionality, by employing for example manifold learning (such as the kPCA \cite{SCH05} or LLE \cite{ROW00}, for instance) or even auto-encoders. The main interest of such a reduction is the possibility of employing regularized regressions, such as the sPGD \cite{sPGD,sPGD-and-variants,parametric-battery}.
\end{remark}

\begin{remark}
	The methodology requires to increase the number of parameters since the boundary conditions of single patches are parametric. However, this does not face the curse of dimensionality contrarily to full-structure based modeling. Indeed, considering a structure composed by $P$ modules each one involving $N_p$ parameters, the total number of parameters in the problem is $PN_p$. Le us split the full problem in $P$ sub-problems involving $N_p$ parameters each and $R$ additional parameters for the interface conditions. Then, the total number of parameters in each local problem is $N_p + R$. Even in the case in which $N_p + R$ is comparable with $PN_p$, the methodology is convenient since all local models are all built in parallel over simple geometries. In target applications involving many modules and many parameters parameters by modules $N_p + R \ll PN_p$ and $R \ll N_p$ ensuring the good scalability of the algorithm.
\end{remark}

\subsection{Computational work-flow for online real-time simulations}

The global proposed workflow consists of:

\begin{enumerate}
	
	\item offline stage:
	
	\begin{enumerate}
		
		\item structure decomposition in a number of modules and determination of the interfaces;
		
		\item creation of a reduced model of the interface conditions: this can be achieved via an a priori parametrization; otherwise, one can compute some high-fidelity solutions of $\mathcal{P}$ for some parameters' combinations and determine the principal modes on the skeleton composed by the interfaces;
		
		\item construction of the parametric solutions for each problem $\mathcal P_p$, for $p=1, \dots , {} P$ (this step is performed efficiently in parallel);
		
	\end{enumerate}
	
	\item online stage:
	
	\begin{enumerate}
		
		\item choosing the model parameters $\vect{p}_1, \dots , \vect{p}_P$;
		
		\item use the part parametric transfer function \eqref{PTF}, $\boldsymbol{u}|_{p} = \boldsymbol{u}|_{p} (\boldsymbol{p}_p, \boldsymbol{\alpha}_s)$ for each module;
		
		\item find the interface parameters ensuring the global equilibrium of modules.
	\end{enumerate}
\end{enumerate}

\begin{remark}
	As previously remarked, during the offline stage, one could perform also the modules assembling by enforcing the interfaces equilibrium. This can be done for many choices of the parameters $\boldsymbol{p}$ to determine the regression $\vect{\Lambda} (\vect{p}_1 , \dots , \vect{p}_P)$. In this case, in the online stage one can directly compute the skeleton kinematics from the already constructed vademecum $\vect{\Lambda} (\vect{p}_1 , \dots , \vect{p}_P)$.
\end{remark}

\section{Results and discussion}
\label{sec:results}

\subsection{Steady state heat conduction}

\begin{figure}[h]
	\centering	
	\includegraphics[scale=0.5]{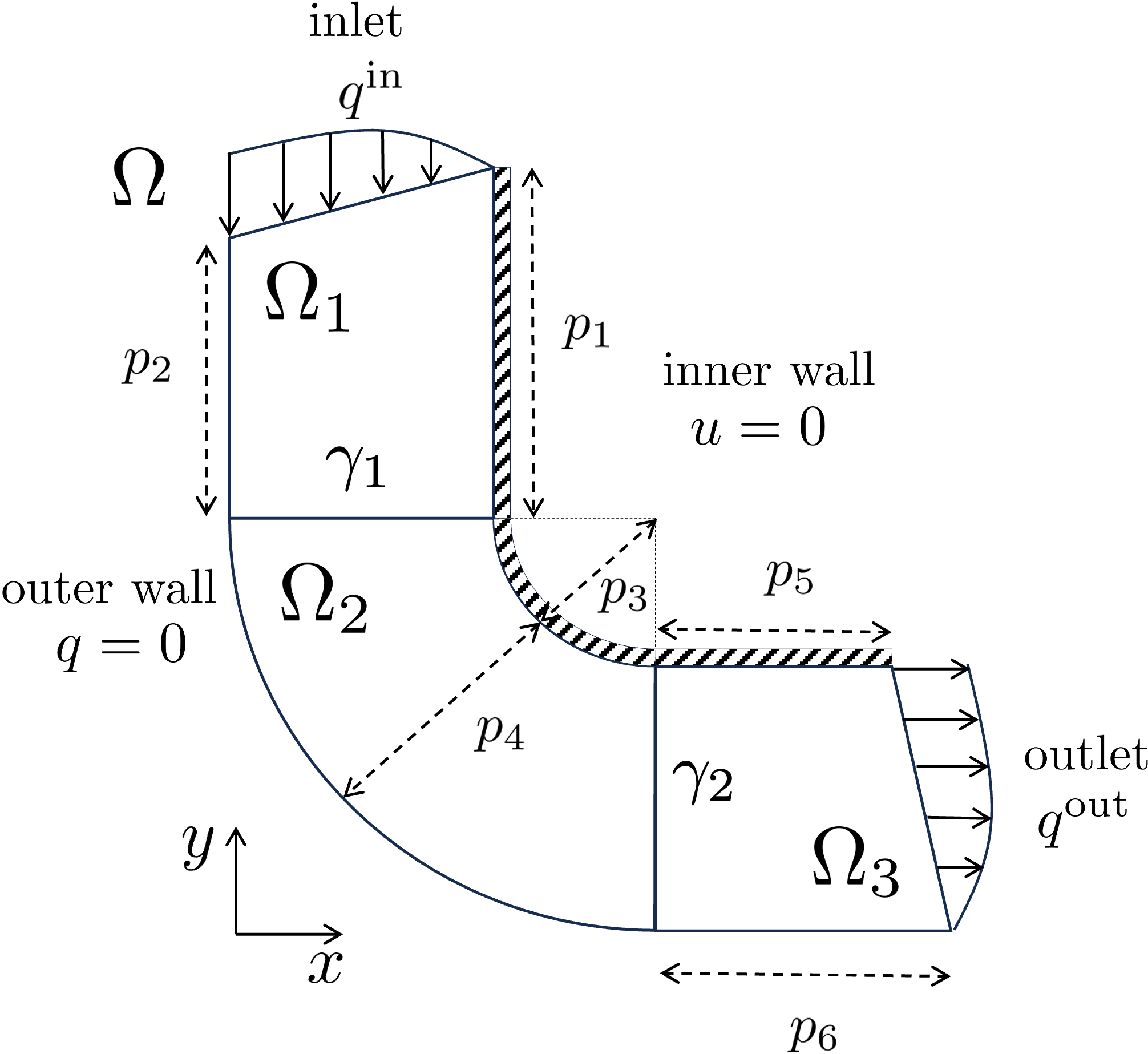}
	\caption{Steady state conduction problem set-up.}
	\label{fig:heat_conduction}
\end{figure}

A first example is a design problem in 2D steady state conduction. The  domain is a curved corner L-shaped geometry having $6$ shape parameters $\boldsymbol{p} = (p_1,\dots,p_6)$, as illustrated in figure \ref{fig:heat_conduction}. Moreover the inlet and outlet fluxes $q_1$ and $q_2$ are parametric and described by 3 coefficients, that is $\boldsymbol{\beta}^{\text{in}} = (\beta_1, \beta_2, \beta_3)$ and $\boldsymbol{\beta}^{\text{out}} = (\beta_4, \beta_5, \beta_6)$, respectively. Null flux and fixed temperature are considered as boundary conditions for the outer and inner wall, respectively. Denoting with $\boldsymbol{\beta} = (\boldsymbol{\beta}^{\text{in}}, \boldsymbol{\beta}^{\text{out}})$ the vector collecting the parameters related to the boundary conditions, the sought parametric solution has 12 parameters and can be written as $u(x,y,\boldsymbol{p}, \boldsymbol{\beta})$, therefore $u$ is a function defined in a 14 dimensions space. 

\begin{figure}[h]
	\centering	
	\includegraphics[scale=0.5]{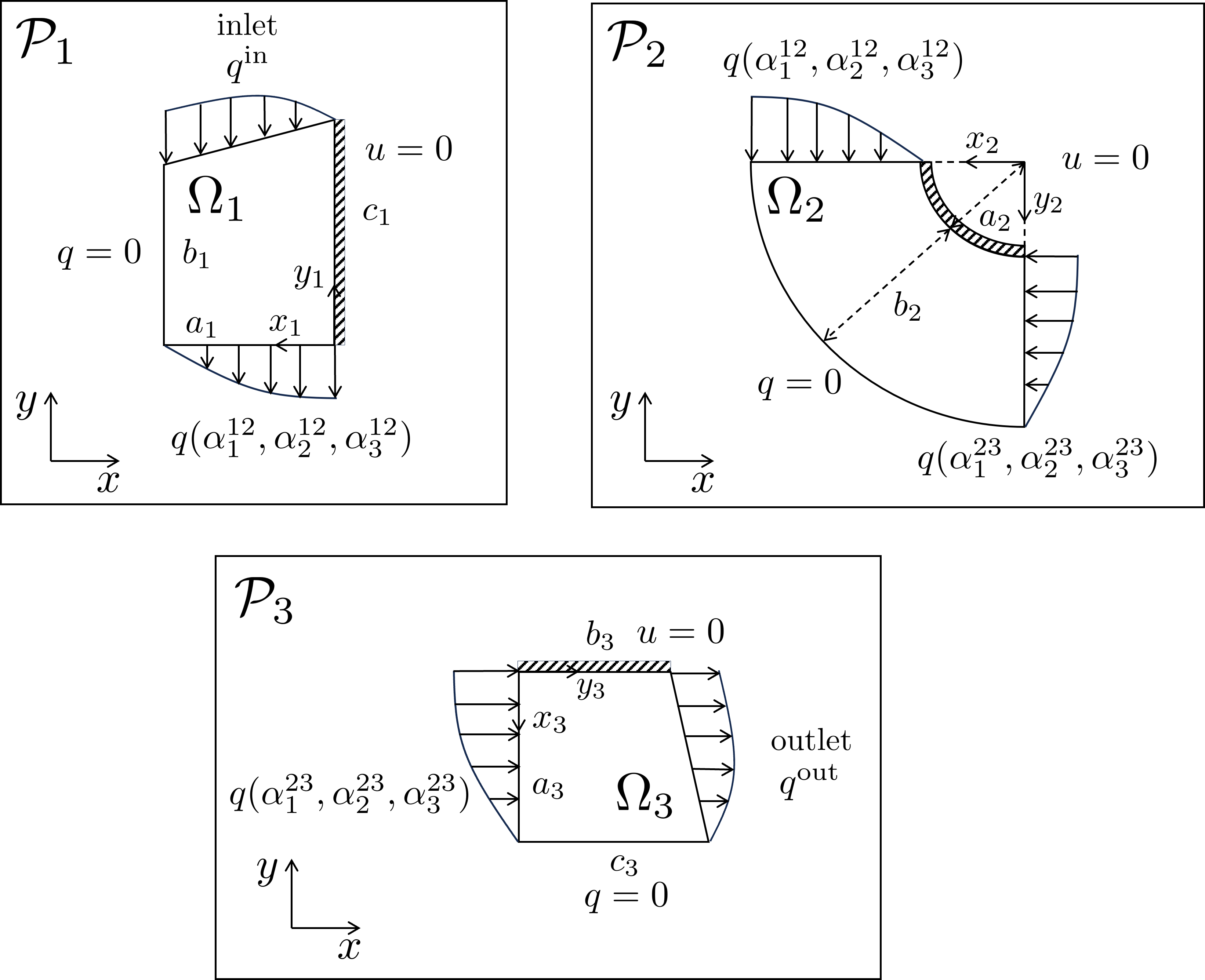}
	\caption{Parametric patches decomposition.}
	\label{fig:heat_conduction_decomposed}
\end{figure}

The problem is decomposed in three parametric sub-problems $\mathcal{P}_i$ defined over $\Omega_i$, for $i = 1, \dots, 3$, having the corresponding geometrical parameters $\boldsymbol{p}_i$. Moreover, the internal interfaces $\gamma_1$ and $\gamma_2$ are treated considering a parametric flux profile still described with 3 parameters, as shown in figure \ref{fig:heat_conduction_decomposed}. For instance, the interface condition between the domain $\Omega_i$ and $\Omega_j$ is an imposed flux depending upon the parameters $\boldsymbol{\alpha}^{ij} = (\alpha_1^{ij}, \alpha_2^{ij}, \alpha_3^{ij})$.

\begin{figure}[h]
	\centering	
	\includegraphics[scale=0.5]{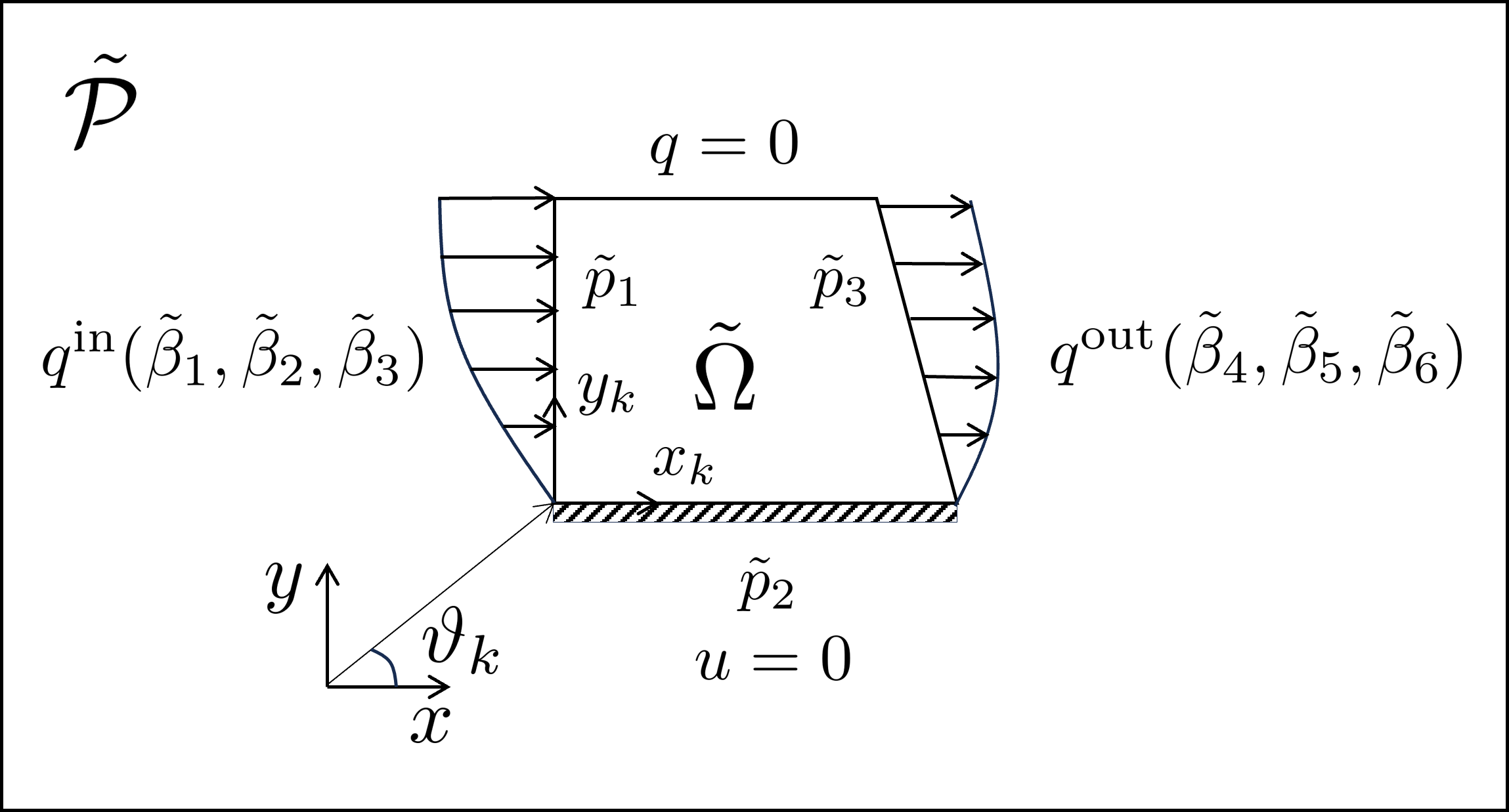}
	\caption{Reference sub-problem for $\mathcal{P}_1$ and $\mathcal{P}_3$.}
	\label{fig:heat_conduction_common_problem}
\end{figure}

Since problems $\mathcal{P}_1$ and $\mathcal{P}_3$ exhibit exactly the same parameters' dependency, they can be reduced to a single parametric problem $\tilde{\mathcal{P}}$ as shown in figure \ref{fig:heat_conduction_common_problem}, where both inlet and outlet fluxes are parametric. The angle $\theta_k$ represents the rotation of the domain to be considered for the assembly within the global system. 

Letting $\tilde{\boldsymbol{p}} = (\tilde p_1, \tilde p_2, \tilde p_3)$ the geometrical parameters and $\tilde{\boldsymbol{\beta}}$ the ones related to the parametric flux, the sought solution of $\tilde{\mathcal{P}}$ is $\tilde u(x,y,\tilde{\boldsymbol{p}}, \tilde{\boldsymbol{\beta})}$ with 9 parameters. The solution $u_i$ of $\mathcal{P}_i$, for $i = \{1, 3\}$, is simply the particularization of $\tilde{u}$ at the correct parameters values, that is $u_i = \tilde{u}(x,y,\boldsymbol{p}_i,\boldsymbol{\beta}_i)$,  where
$$
\begin{cases}
	\boldsymbol{p}_1 = (a_1, c_1, b_1), \\
	\boldsymbol{\beta}_1 = (\boldsymbol{\beta}^{\text{in}}, \boldsymbol{\alpha}^{12}), 
\end{cases}\quad \text{and} \quad
\begin{cases}
	\boldsymbol{p}_3 = (a_3, b_3, c_3), \\
	\boldsymbol{\beta}_3 = (\boldsymbol{\alpha}^{23}, \boldsymbol{\beta}^{\text{out}}). 
	\end{cases}
$$

In the same way, the solution of problem $\mathcal{P}_2$ has 2 geometrical parameters $\boldsymbol{p}_2 = (a_2, b_2)$ and $6$ parameters for the inlet and outlet fluxes, that is $\boldsymbol{\alpha}^{12}$ and $\boldsymbol{\alpha}^{23}$, respectively.

The three parametric sub-solutions finally read
\begin{equation}
	\label{eq:subsolutions}
	\begin{cases}
		u_1(x,y,\boldsymbol{p}_1,\boldsymbol{\beta}^{\text{in}}, \boldsymbol{\alpha}^{12}),\\
		u_2(x,y,\boldsymbol{p}_2,\boldsymbol{\alpha}^{12}, \boldsymbol{\alpha}^{23}),\\
		u_3(x,y,\boldsymbol{p}_3, \boldsymbol{\alpha}^{23}, \boldsymbol{\beta}^{\text{out}}),
	\end{cases}
\end{equation}
and the global solution $u(x,y,\boldsymbol{p},\boldsymbol{\beta})$ is obtained by assembly. To this purpose, one must ensure the geometric constraint fixing $a_1 = b_2 = c_3 = p_4$ and, for a given choice of parameters, minimize the temperature jumps at the interface, by solving
\begin{equation}
	\displaystyle \min_{\boldsymbol{\alpha}^{12},\boldsymbol{\alpha}^{23}} (\norm{u_1 - u_2}_{2,\gamma_1} + \norm{u_2 - u_3}_{2,\gamma_2}),
\end{equation}
where $\norm{\cdot}_{2,\gamma_{i}}$ denotes the Euclidean norm at the interface $\gamma_{i}$.

In this example, the geometrical parameters are chosen varying in $[1,3]$ while all the coefficients of flux profiles have range $[-100,100]$. The metamodels $\tilde{u}$ (i.e., $u_1$ and $u_3$) and $u_2$ are computed in parallel, employing the NURBS-PGD method. Each direction (space and parameters ones) is discretized in $51$ nodes, yielding a total number of DOFs of $51^{11}$ and $51^{10}$ for $\tilde{\mathcal{P}}$ and $\mathcal{P}_2$, respectively. This is possible thanks to the usage of PGD-based separated representations.

\begin{figure}[h]
	\centering	
	\includegraphics[scale=0.2]{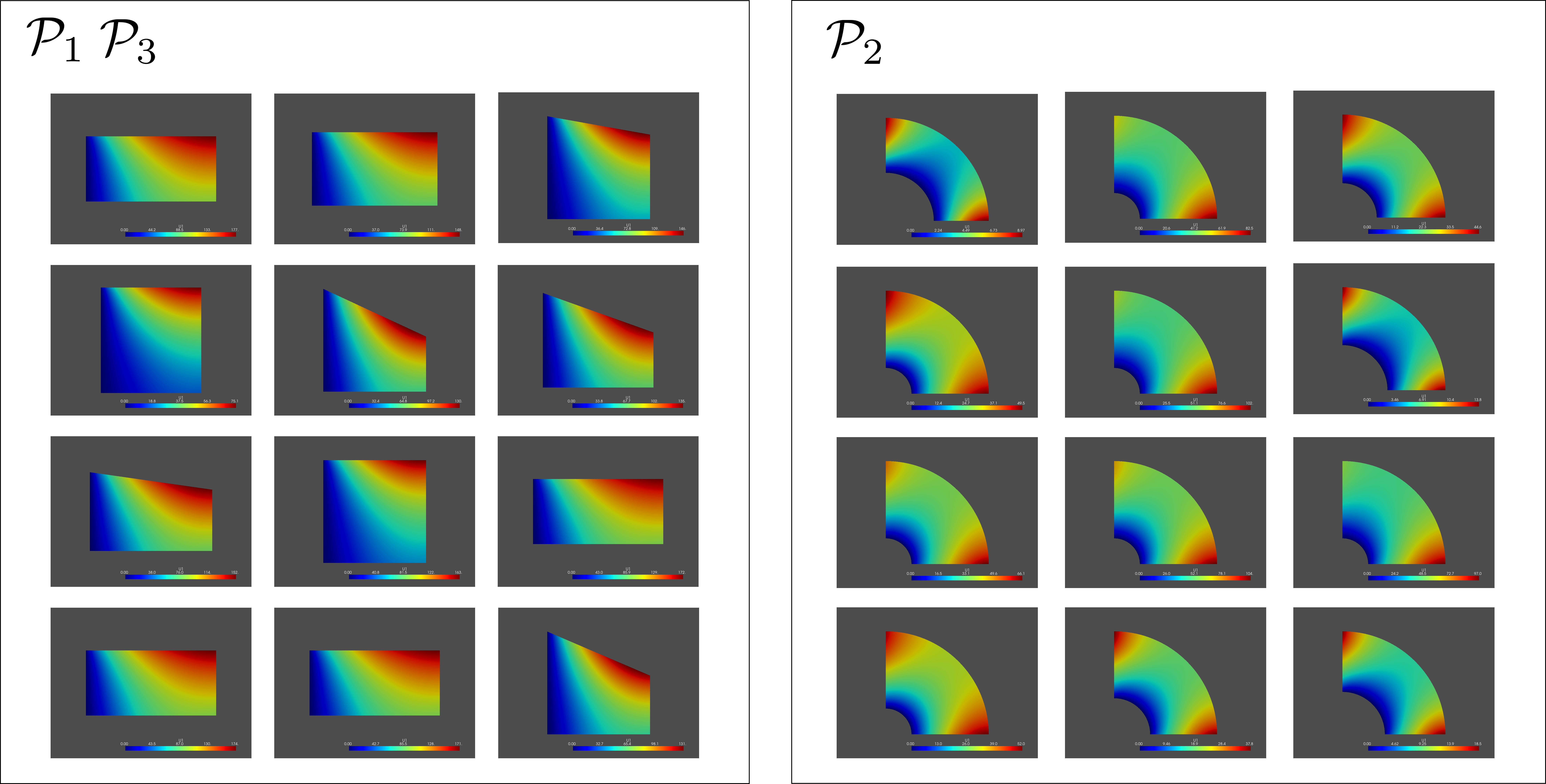}
	\caption{12 snapshots of parametric sub-solutions.}
	\label{fig:heat_conduction_subproblems_results}
\end{figure}

Figure \ref{fig:heat_conduction_subproblems_results} shows 12 snapshots of the sub-solutions for different combinations of the parameters. This can be seen as a catalog of patches which can suitably be combined in the global assembly to evaluate several designs of the curved $L$-shape geometry of figure \ref{fig:heat_conduction}.

\begin{figure}[H]
	\centering	
	\includegraphics[scale=0.2]{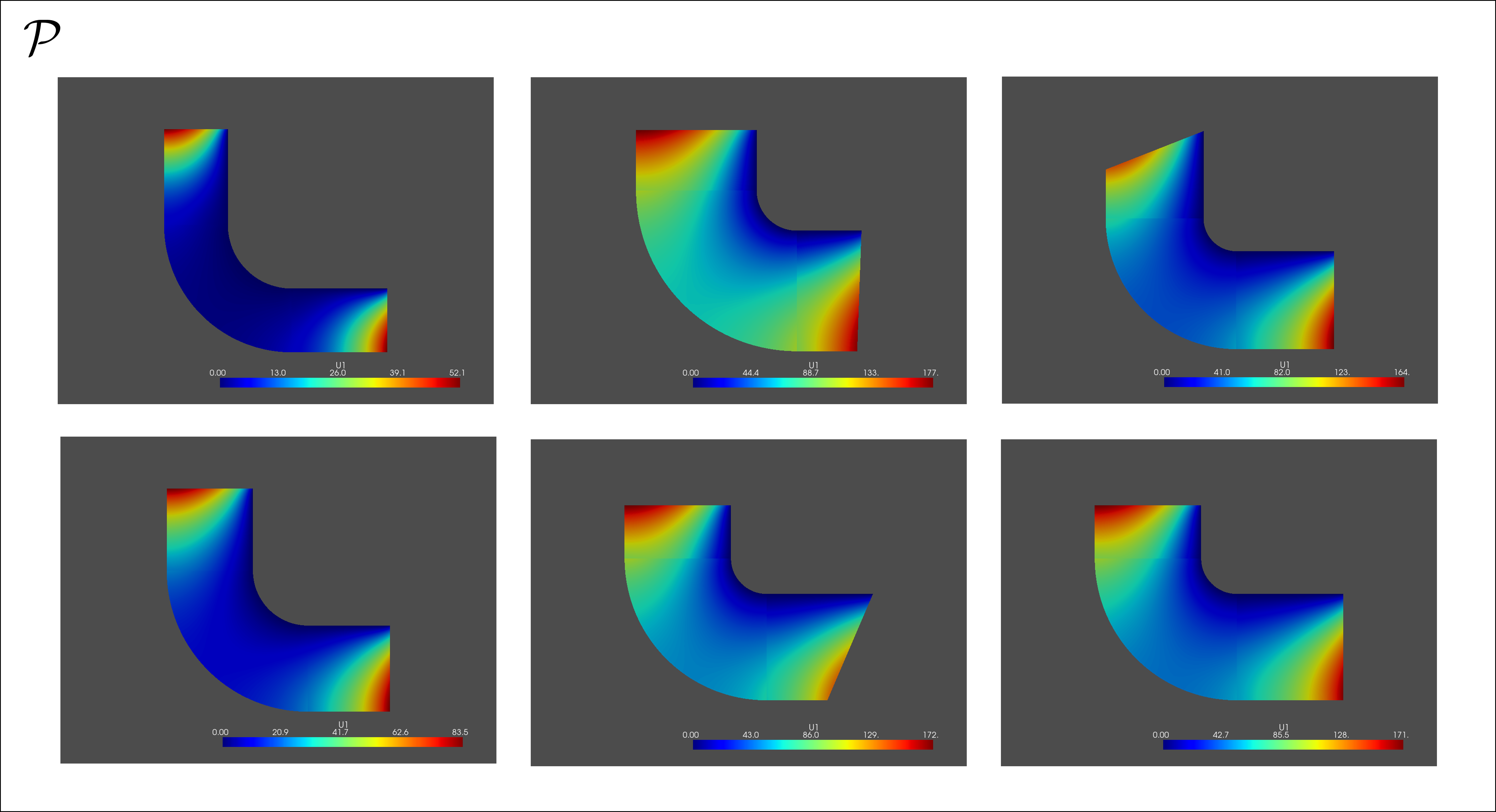}
	\caption{6 snapshots of the assembled parametric solution.}
	\label{fig:heat_conduction_problem_results}
\end{figure}

\newpage
\subsection{First order plate deformation}

\begin{figure}[h]
	\centering	
	\includegraphics[scale=0.5]{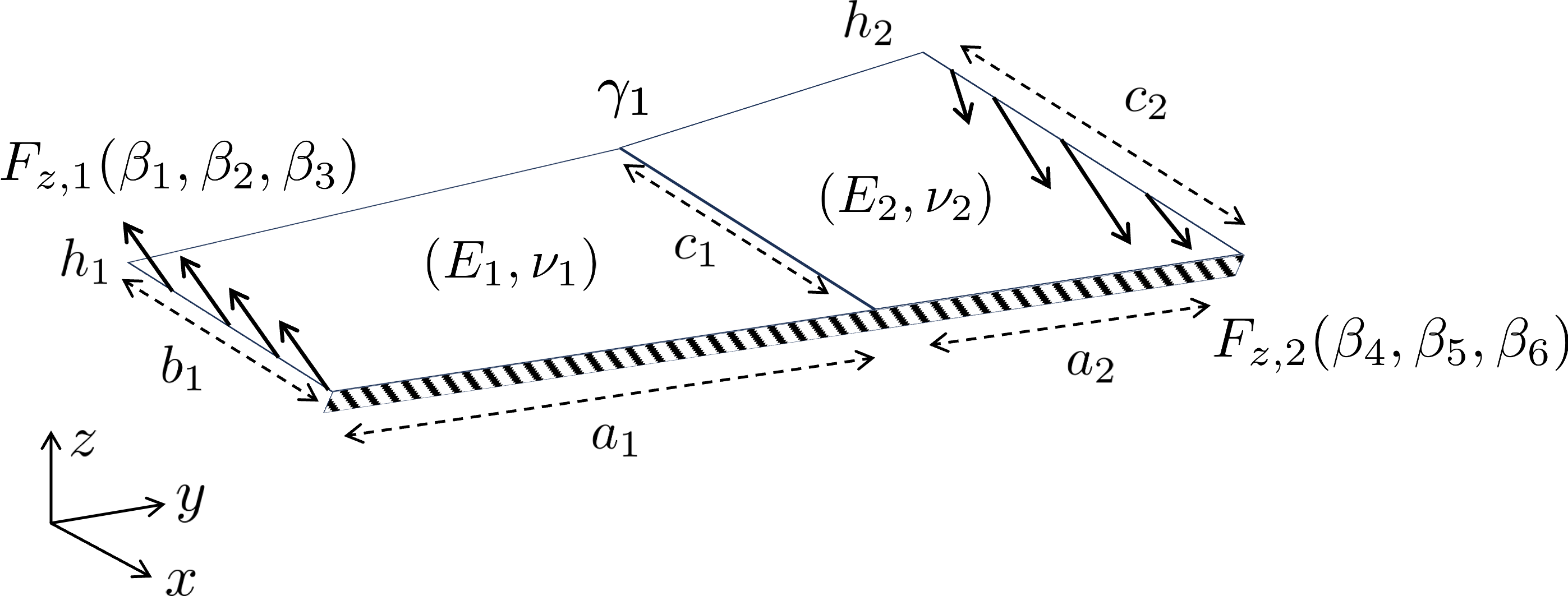}
	\caption{Plate problem set-up.}
	\label{fig:plate}
\end{figure}

Let us consider a thin elastic plate composed of two modules with different geometry and material properties, as shown in figure \ref{fig:plate}. The plate is clamped along one edge (homogeneous Dirichlet boundary condition), while along the left and right edges a distributed out-of-plane loading is applied. The upper edge has a traction-free boundary condition (homogeneous Neumann). The mechanical problem $\mathcal{P}$ has 17 geometric and model parameters resumed in table, with also the corresponding ranges \ref{tab:parameters-plate}.

\begin{table}[h]
	\small
	\centering
	\begin{tabular}{|c|c|c|c|c|}
		\hline
		Length                   & Thickness     & Young Modulus           & Poisson Ratio  & Force       \\ \hline
		$a_1, b_1, c_1, a_2, c_2$ & $h_1, h_2$      & $E_1, E_2$              & $\nu_1, \nu_2$ & $\beta_1, \dots, \beta_6$ \\ \hline
		$[0.1, 0.3]$              & $[0.001, 0.01]$ & $[100, 300] \cdot 10^9$ & $0.2, 0.4$     & $[-1,-1] \cdot 10^4$      \\ \hline
	\end{tabular}
	\caption{Problem parameters and ranges.}
	\label{tab:parameters-plate}
\end{table}

Each out-of-plane force $F_{z,i}$, for $i = {1, 2}$ is assumed depending upon $3$ coefficients $\boldsymbol{\beta}_i$. The two plates share the interface $\gamma_1$ whose length is determined by the parameter $c_1$. 

The plate is modeled through the first order Kirchhoff theory, which expresses the displacement field in terms of the middle plane kinematic variables $w$, $\theta_x$ and $\theta_y$, 
\begin{equation}
	\left\{
	\begin{aligned}
		u(x,y,z) & = z \theta_y (x, y) \\
		v(x,y,z) & = -z \theta_x (x, y) \\
		w(x,y,z) & = w (x,y)
	\end{aligned}
	\right.
\end{equation}
where $\theta_x$ and $\theta_y$ are the angles defining the rotation of the normal vector to the middle surface and $w$ is the vertical displacement (deflection).

\begin{figure}[h]
	\centering	
	\includegraphics[scale=0.5]{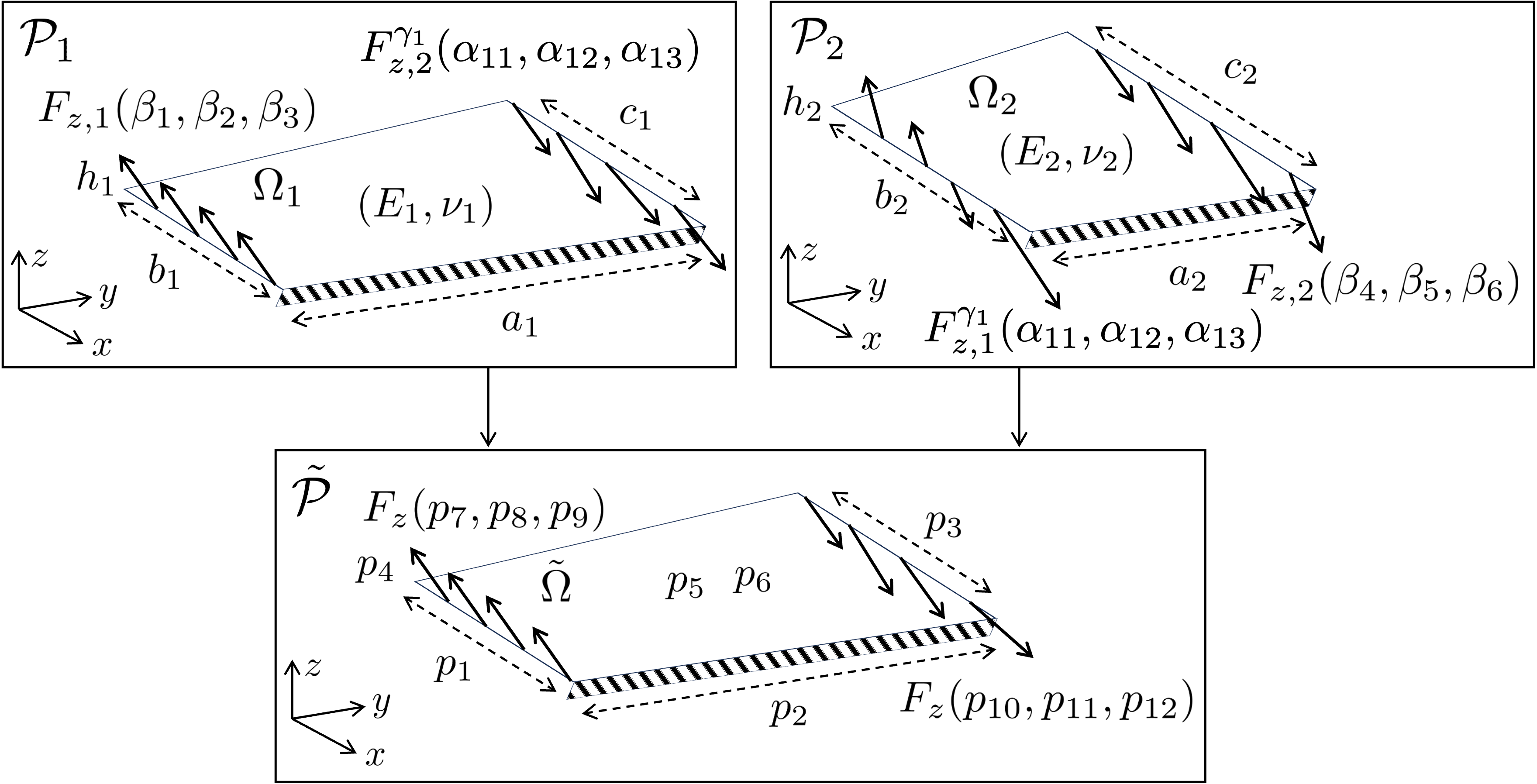}
	\caption{From modules to a reference parametric patch.}
	\label{fig:plate_patch}
\end{figure}

Following the same procedure, problem $\mathcal{P}$ can be split in two sub-problems which can be reduced to the same parametric module $\tilde{\Omega}$, as resumed in figure \ref{fig:plate_patch}.

In this way, the original problem is reduced to a single problem $\tilde{\mathcal{P}}$ having $12$ parameters, and the two particularized sub-solutions can be written as
\begin{equation}
	\label{eq:subsolutions-plate}
	\begin{cases}
		\boldsymbol{u}_1(x,y,z,\boldsymbol{p}_1,\boldsymbol{\beta}_1,\boldsymbol{\alpha}^{12}),\\
		\boldsymbol{u}_2(x,y,z,\boldsymbol{p}_2,\boldsymbol{\alpha}^{12}, \boldsymbol{\beta}_2),
	\end{cases}
\end{equation}
where $\boldsymbol{p}_i = (a_i, b_i, c_i, h_i, E_i, \nu_i)$ for $i = \{1, 2\}$.

Finally, the assembly is performed imposing the geometrical constraint $c_1 = b_2 = c_3$ and the minimization of the displacement jump at the interface, that is
\begin{equation}
	\displaystyle \min_{\boldsymbol{\alpha}^{12}} \norm{\boldsymbol{u}_1 - \boldsymbol{u}_2}_{2,\gamma_1},
\end{equation}
where $\norm{\cdot}_{2,\gamma_1}$ denotes the Euclidean norm at the interface $\gamma_1$.

\begin{figure}[H]
	\centering	
	\includegraphics[scale=0.5]{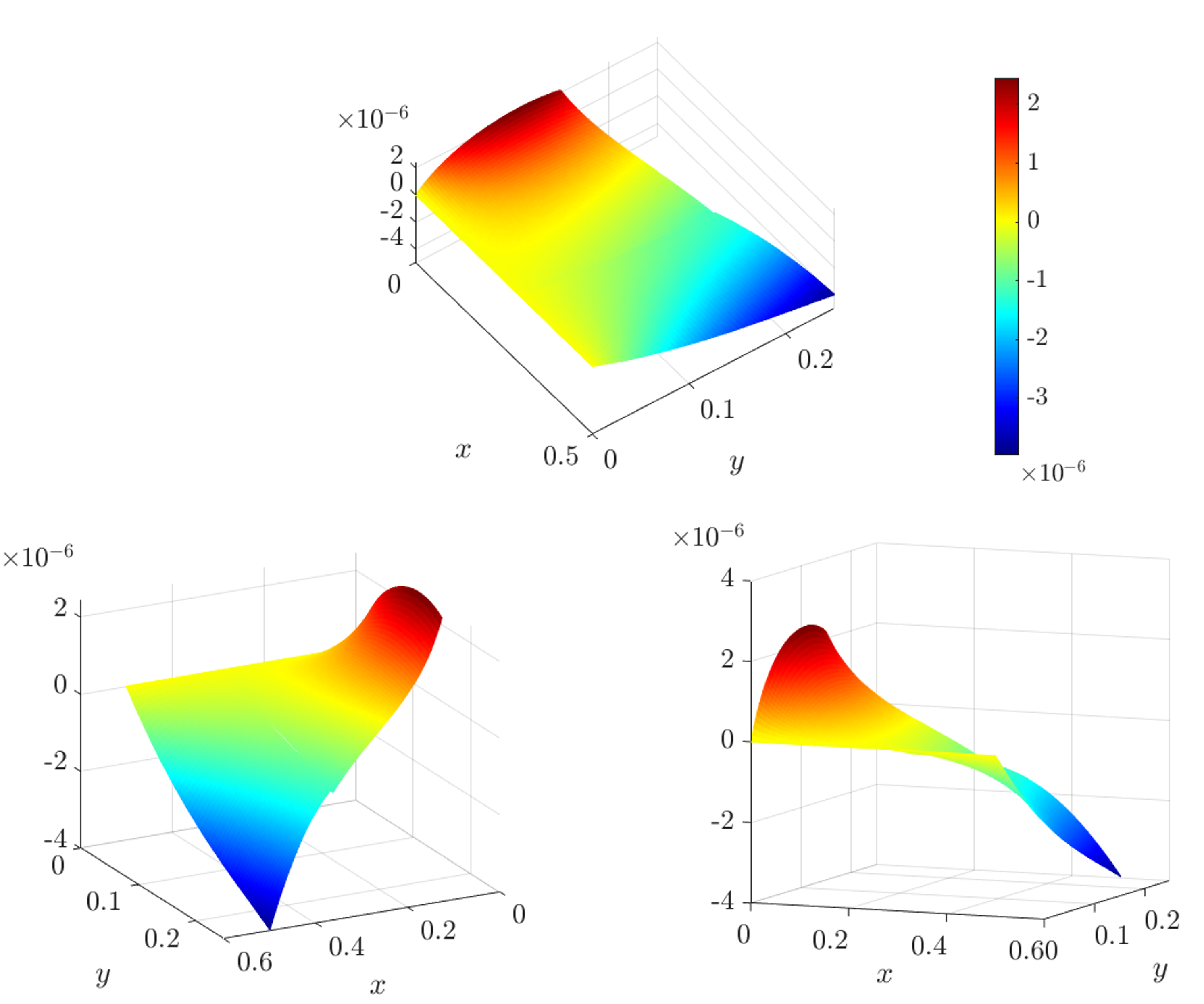}
	\caption{Plate bending from different isometric views.}
	\label{fig:plate_problem_results}
\end{figure}

For instance, figure \ref{fig:plate_problem_results} shows the assembled solution when choosing the following parameters values
\begin{equation}
	\begin{cases}
		\boldsymbol{p}_1 = (0.3, 0.15, 0.2, 0.01, 100\cdot 10^9, 0.3),\ \boldsymbol{\beta}_1 = (7\cdot 10^3,7\cdot 10^3,7\cdot 10^3), \\
		\boldsymbol{p}_2 = (0.2, 0.2, 0.25, 0.01, 200\cdot 10^9, 0.3),\ \boldsymbol{\beta}_2 = (0,-7\cdot 10^3,-7\cdot 10^3).
	\end{cases}
\end{equation}

\section{Conclusions}
\label{sec:conclusions}

In this work a novel methodology to build parametric surrogates in high-dimensional problems has been proposed. 

The method is based on a modularization of the domain in non-overlapping parametric modules which are treated separately and simultaneously (in the sense of parallel computing). The arising sub-solutions can be assembled in the global system by means of a suitable minimization ensuring the compatibility at the modules interfaces. 

The replicability of the modules is the most appealing feature of the technique, which can be exploited for the quick design and performance evaluation. This may be useful in many industrial and large-scale applications, which are being addressed in our current research. In particular, special care is paid to the minimization step, which is fundamental to ensure a real-time response.

\section*{Declarations}

\subsection*{Availability of data and material} Interested reader can contact authors.
	
\subsection*{Competing interests}
The authors declare that they have no competing interests.
	
\subsection*{Author's contributions}
All the authors participated in the definition of techniques and algorithms.  
	
\subsection*{Acknowledgements}
Authors acknowledge the support of the ESI Group through its research chair CREATE-ID at Arts et Métiers ParisTech.
	
This research is part of the program DesCartes and is supported by the National Research Foundation, Prime Minister’s Office, Singapore under its Campus for Research Excellence and Technological Enterprise (CREATE) program.

\bibliography{references}



\renewcommand{\thesubsection}{\Alph{subsection}}

\end{document}